\documentclass[prb,twocolumn,showpacs,preprintnumbers,amsmath,amssymb,floatfix]{revtex4}
\usepackage{graphicx}
\usepackage{dcolumn}
\usepackage{bm}
\usepackage{longtable}
\begin{document}

\title{ Strain induced metal-insulator transition in ultrathin films of SrRuO$_3$} 
\author{Kapil Gupta\footnote{Present address: Indo-Korea Science and Technology Centre, Bangalore - 560029 and Jawaharlal Nehru Centre for Advanced Scientific Research, Bangalore - 560064}} 
\affiliation{Department of Condensed Matter Physics and Materials Science, 
S.N. Bose National Centre for Basic Sciences, JD Block, Sector III, Salt Lake, Kolkata 700098, India }
\author{Basudeb Mandal and Priya Mahadevan\footnote{Corresponding Author: priya@bose.res.in}}
\affiliation{Department of Condensed Matter Physics and Materials Science, 
S.N. Bose National Centre for Basic Sciences, JD Block, Sector III, Salt Lake, Kolkata 700098, India }
\date{\today}

\begin{abstract}
The ultrathin film limit has been shown to be a rich playground for unusual low dimensional physics. 
Taking the example of SrRuO$_3$ which is ferromagnetic and metallic at the bulk limit, one finds that 
it becomes antiferromagnetic and insulating at the three monolayers limit when grown on SrTiO$_3$. The 
origin of the insulating state is traced to strongly orbital dependent exchange splittings.
A modest compressive strain of 1\% of the SrTiO$_3$ substrate
is then found to drive the system 
into a highly confined two-dimensional 100\% spin polarized metallic state. This metal-insulator transition 
driven by a modest strain could be useful in two state device applications. 
\end{abstract}

\maketitle

\section {Introduction}
The birth of modern day electronics began with semiconductor
technology. However, as device dimensions are reaching limits
where their operation is no longer feasible without losses, 
alternate materials are being investigated for new generation
electronics. Transition metal oxides are one such class of
materials being explored as possible candidates\cite{ohtomo,ohtomo1}.  In contrast
to semiconductor heterostructures, here, the strongly
coupled spin, charge and lattice degrees of freedom lead to 
very diverse phenomena even with small deviations in the 
parameter space. One such parameter that has been used to tune 
the properties of transition metal oxides is strain\cite{strain,strain1,strain2,strain3}, where
in some instances one has been able to render nonmagnetic
materials ferromagnetic\cite{fuchs,fuchs1,kapil}, in certain others one is able to use
strain to induce ferroelectricity\cite{satadeep,rabe1,rabe2,rabe3} and so on. Another key 
parameter that controls the properties of the films has been
the choice of the substrate. This can be used to tune a
different crystal structure for the films grown on top than
is usually favored\cite{zayek}. The films then adopt the new crystal structure
for few nanometers till one has strain relaxation that takes it
to the crystal structure favored in the bulk.

In this work we consider the example of SrRuO$_3$. This is both
metallic and ferromagnetic in the bulk. Since SrRuO$_3$ involves
a 4$d$ transition metal atom, which have wide bands, the expectation was that when
ultrathin films were grown on a substrate, it would retain
its metallicity down to the ultrathin limit. However, it was
shown experimentally that below four monolayers of SrRuO$_3$, the
system was insulating\cite{toyota,xia}. First principle electronic
structure calculations were found to support this view and showed
that lattice distortions drove the insulating state at the three
monolayers limit\cite{priya}. In this work we consider the three monolayers
limit and examine if one
can retain metallicity and stop the metal to insulator transition
by subjecting the films to compressive strain.
This was indeed found to be the case and a modest 
compressive strain of 1\% was found to be sufficient.
The metallic state at the three monolayers limit 
was found to be highly confined in two
dimensions and was found to be completely spin polarized , similar
to what has been suggested in sandwich structures of SrTiO$_3$
and SrRuO$_3$\cite{ghos}. The
insulating state obtained in the absence of any strain however, 
was found to have a surprising origin. The lattice distortions
of the RuO$_6$ octahedra result in a level ordering in which the
$d_{xz}$, $d_{yz}$ orbitals are at a lower energy compared to
the $d_{xy}$ orbitals. Indeed we find such a level ordering
in the majority spin channel when we examine the density of states. However, one finds a reversal of the 
level ordering in the down spin channel. This is traced to
the differences in the exchange splitting between the $d_{xy}$
and the $d_{xz}$/$d_{yz}$ orbitals which arises from the superlattice geometry that one has in which
the $d_{xy}$ orbitals have wider bands associated with them than the
$d_{xz}$ and $d_{yz}$ orbitals. Under compressive strain
one can change the relative contributions of the energy gain arising
from hopping with respect to that from
the intraatomic exchange interaction. This can be used to control 
which orbital is
occupied in the minority spin channel. This
then has been used to bring a crossover to a spin polarized 
metallic state with the fourth electron occupying the $d_{xz}$
and $d_{yz}$ levels. In contrast to the work 
by Verrisimo-Alves {\it et al.}\cite{ghos} who find the highly confined
two dimensional 100\% spin polarised electron gas in
superlattices of SrRuO$_3$/SrTiO$_3$, we find this effect
with just one monolayer of SrO on top of the RuO$_2$ layer
{\it i.e} the three monolayers limit. So this demonstrates
that the ultrathin limit serves as a playground for 
manipulating various atomic interaction strengths and allows one to arrive
at unusual aspects of the electronic structure
which are not found in the bulk limit.

\section {Methodology}

The electronic structure of bulk as well as thin films of SrRuO$_3$ was
calculated within a planewave pseudopotential
implementation of density functional theory within Vienna ab-initio simulation package (VASP)\cite{vasp,vasp1}. 
The GGA (Generalized Gradient Approximation)-PBE (Perdew–Burke–Ernzerhof) approximation
to the exchange correlation functional\cite{gga} was used. Correlation effects on Ru
were treated within the GGA+U method using the formalism of Dudarev 
\cite{dudarev}. A value of U=2.5 eV and J=0.4 eV was
applied on the Ru atom as deduced from the constrained random phase approximation\cite{rpa} based formalism. 
Inspite of the results being calculated from a first principles estimate of U, we have varied U as well as
the double counting scheme used to illustrate the sensitivity of the 
results to the choice of $U$. However, the constrained RPA determined 
value of $U$ is able to reproduce various limits observed experimentally
indicating the predictive power of the approach.
A k-point mesh of $6\times6\times6$ and $6\times6\times2$ was used for
the bulk and thin film calculations
respectively.
It was increased to $8\times8\times8$ and $8\times8\times2$ to calculate
the density of states. In addition an energy cutoff of 400 eV was used for the kinetic energy of the plane waves included in the basis.
Spheres of radii equal to 0.9 \AA~ were used to calculate the 
 Ru d projected partial density of states.

The experimental structure was taken for bulk SrRuO$_3$\cite{expt} and the internal coordinates were
optimized. In order to calculate the electronic structure of the ultrathin films of SrRuO$_3$,
we considered a symmetric slab consisting of 15 layers of TiO$_2$ and SrO 
growing in the $(001)$
direction. The in-plane lattice constant was kept fixed at the experimental lattice constant of SrTiO$_3$ which is 3.905 \AA. 
This is smaller than the pseudocubic lattice constant of SrRuO$_3$ found to be 
3.92 \AA. The substrate lattice constant was varied to simulate the 
effects of strain. The substrate
was taken to terminate with the TiO$_2$ surface on which SrO/RuO$_2$ layers were added alternately.
A vacuum of 15 \AA~ was used to minimize the interaction between images of the slab. As GdFeO$_3$
type of distortions are found in bulk SrRuO$_3$, we allowed for both rotations as well as tilts of the octahedra.
Again, as in the case of the bulk calculations, here also the internal coordinates were optimized.
Lattice mismatch with the substrate imposes a compressive strain of 0.4\% on SrRuO$_3$ thin films. 
These films were also considered on 1 and 2\% compressed SrTiO$_3$, which leads to 1.4 and 2.4\% compressive 
strain on the thin films of SrRuO$_3$. 

{\section {Results and Discussion}}

{\subsection {Bulk SrRuO$_3$}}
SrRuO$_3$ is found to be ferromagnetic and metallic in the bulk and
favors an orthorhombic unit cell. The orthorhombicity is driven by
both GdFeO$_3$ rotations of the RuO$_6$ octahedra as well as the tilts\cite{low}.
Before we examine the properties of SrRuO$_3$ in the thin film form, we
first examine the bulk structure in our calculations. The ferromagnetic
metallic unit cell is found to be the ground state in our calculations.
Comparing the structural parameters of our optimized structure with
experiment, we find that the calculations get the Ru-O bondlengths
in reasonable agreement with experiment. The bond angles are found
to be 158$^{\circ}$ in the ac plane, slightly underestimated from the experimental
values which are found to be in the range 161 - 163$^{\circ}$ as shown in Table I. The 
bond angles in the b-direction are found to be underestimated by 3-5$^{\circ}$
from the experimental values\cite{expt,expt2,expt3}.

{\subsection {Two monolayers of SrRuO$_3$ on SrTiO$_3$}}
As discussed in the Methodology section, two or more monolayers of SrRuO$_3$ are
grown on SrTiO$_3$. The rotations of the succesive octahedra
stacked in the $c$-direction in SrTiO$_3$ are out of phase and this has been included in the calculations.
Additionally one finds that the substrate imposes a tetragonal
crystal structure on the SrRuO$_3$ overlayers. We first examine
the case where we have two monolayers of SrRuO$_3$ grown on TiO$_2$
terminated SrTiO$_3$ substrates. Photoemission experiments
indicate that these films are insulating, \cite{ghos} and our calculations
also find them to be so. In order to examine the origin
of the insulating state in our calculations, we examine the
distortions of the RuO$_5$ motifs in our optimized unit cell. 
These are shown in Fig. 1. The distortions in the $ab$ plane are
found to consist of Ru-O bondlengths equal to 1.95 and 1.97 \AA~
as shown in Fig. 1(a). The out-of-plane Ru-O bondlength is
found to be 2.15 \AA, dramatically modified from the in-plane
Ru-O bondlengths. This suggests that the surface RuO$_2$ layer
is weakly coupled to the substrate. Further the Ru environment
is found to approach a square planar geometry. The large 
structural distortions observed here for the RuO$_5$ motifs
would involve a large energy cost in terms of the strain energy in
increasing the length of the Ru-O bond in the 
c-direction. So, the natural question is to 
understand where the energy of this distortion is coming from 
and why it is happening in the first place. Ru in SrRuO$_3$ has a 
$d^4$ configuration. In early 3d transition metal oxides one has a smaller crystal
field splitting than the exchange splitting. However for 4d oxides,
one has a larger exchange splitting than the crystal field splitting.
This results in the Ru d states with $t_{2g}$ symmetry being completely
filled in the up spin channel and the fourth electron goes into the $t_{2g}$ down spin
channel. This is the energy level diagram for bulk SrRuO$_3$. At the two monolayers
limit one has seen earlier\cite{priya} 
that the symmetry about the Ru site is reduced 
to square pyramidal. The level ordering is
dictated by the long Ru-O bond in the z-direction. We have the $t_{2g}$ levels splitting 
into $d_{xz}$ and $d_{yz}$ at lower energies compared to the $d_{xy}$ orbital. The orbitals
with $e_g$ symmetry split into the the $d_{z^2}$ orbital at lower energy compared to the $d_{x^2-y^2}$
orbital. As a result we have the four electrons on Ru occupying the majority $t_{2g}$-derived orbitals
and then the $d_{z^2}$ orbital. Hence we have a rare occurrence of a high spin state at
the Ru site. The gain in energy from the spin-state transition also explains why one can sustain
the long Ru-O bond in the z-direction.

Another puzzling aspect that we find is the polar nature of the distortions
of the Ru-O bonds in the ab-plane. This probably arises from the fact that 
the surface distortions have driven the system into a band insulator. 
The system can have weak second order Jahn-Teller effects and this is 
what we find here.
The Ru atom is found to offcentre towards a pair of oxygens in the ab-plane
and as a result a pair of oxygens have shorter Ru-O bondlengths of 1.95 \AA~
than the other two (1.97 \AA). The magnitudes of these distortions 
decrease when we include the tilts of the octahedra\cite{hatt}. Additionally we 
find that the net electric polarization is zero as the dipole moments
associated with different RuO$_5$ motifs are oriented in opposite directions. 
As discussed earlier, the Ru-O-Ru angles for bulk SrRuO$_3$ are found to be
158$^{\circ}$ for the in-plane case and 160$^{\circ}$ for the out of plane case. In the present
case we find the bond angles equal to 167$^{\circ}$ and 170$^{\circ}$.
These deviations in the bond angles as large as 8$^{\circ}$-10$^{\circ}$ 
from the values found for bulk SrRuO$_3$ are surprising, especially
since compressive strain due to the substrate should result in shorter
bonds and a more distorted Ru-O network. These expectations are based on
our notional understanding of the origin of
GdFeO$_3$ distortions. A smaller ion at the A site in a perovskite 
lattice of the form ABO$_3$,
results in a smaller volume for the perovskite. This also leads to shorter bonds between the transition
metal, B, and oxygen, which increases the repulsion between the electrons on B and oxygen. The structure,
then distorts with the BO$_6$ octahedra rotating. This distortion, known as 
GdFeO$_3$ distortion is commonly observed in perovskite oxides, and leads to smaller B-O-B angles in the 
perovskite oxides with unit cell of smaller volume. 
The compressive strain of the substrate is
expected to behave similarly.
Contrary to these expectations, one instead finds an increase here.
This could possibly arise from an attempt by
the system to increase its bandwidth, as the effectively square planar geometry that is favored leads to a 
further loss of bandwidth than linked RuO$_5$ motifs in the z-direction.

{\subsection {Three monolayers of SrRuO$_3$ on SrTiO$_3$}}
Adding a layer of SrO on the two monolayers of SrRuO$_3$ results
in the in-plane Ru-O network to adopt the structure shown in Fig. 2. 
Each Ru atom has a small 
Jahn-Teller distortion with the long and short Ru-O bonds 
differing by 0.01 \AA. The in-plane
Ru-O-Ru angles now at this limit of three monolayers are found to be 152$^\circ$,
6-8$^\circ$ less than the values found
in bulk. This trend, however is expected in the case of compressive strain
as discussed earlier. The out-of-plane
bondlengths are found to be 2.0 \AA \space 
and 2.05 \AA. The longer Ru-O 
bondlength in the z-direction
results in a degeneracy lifting of the t$_{2g}$ orbitals with 
the d$_{xz}$ and d$_{yz}$ levels
found at lower energies compared to the d$_{xy}$ orbitals, as 
seen for the Ru d projected
partial density of states for the up spin channel in
Fig. 3(a)-(e). However one finds a change in the level ordering in 
the down spin channel. The fourth
electron goes into the down spin d$_{xy}$ orbital. This could be 
understood in terms of an 
orbital dependent exchange splitting, the origin of which can be traced back to the itineracy of the
electron in the different d orbitals.
The electron in d$_{xy}$ orbitals delocalize in the xy-plane
forming wide bands, while those in the d$_{xz}$ and d$_{yz}$ orbitals couple via hopping with other d$_{xz}$ and d$_{yz}$
orbitals only along the x- and y-axis respectively and form narrower bands. 
The hopping in the z-direction is very weak,
as the corresponding Ti orbitals to which they can hop to are much higher in energy.
As a result the exchange splitting for the d$_{xy}$ orbitals
is smaller than that of the d$_{yz}$ and d$_{xz}$ orbitals and hence the former gets occupied.
This is shown schematically in Fig 5(a). 
Thus the electronic structure brings out unusual aspects of the physics of this regime and enables us
to manipulate interactions at the atomic level.

We then went on to examine whether the system would remain insulating under additional compressive strain.
This was simulated by considering the compressed lattice parameter of the SrTiO$_3$ substrate, and
subjecting it to 1\% and 2\% compressive strain. 
Considering the 1\% strained 
case, we find the in-plane bondlengths to be 1.98 and 2.0 $\AA$ after relaxation, while the
out of plane bondlengths are found to be 2.07 and 2.04 \AA~along negative and positive z-direction respectively.
The in-plane 
bond angle is found to be 152.2$^{\circ}$, smaller than the bulk value as expected. 
Examining the density of states (Fig. 4(a)-(e)), 
we find that the level ordering in the majority spin channel is the same 
as when the substrate was unstrained, and we have $d_{yz}$ and $d_{xz}$ 
orbitals at lower energies compared to the $d_{xy}$ orbital. The same
level ordering is found in the down spin channel also and this is shown schematically in Fig 5-(b).
This arises from the shorter Ru-O bonds that one has in the present case, 
which result in larger p-d hopping interaction strengths. Hence in the
minority spin channel, the $d_{xy}$ levels remain above the $d_{xz}$
and $d_{yz}$ levels. This results in a metallic ground state \cite{paramag}. For 2\% compressed SrTiO$_3$
substrate, structure and density of states remain the same qualitatively. In this case in-plane bondlengths are 
found to be 1.99 and 2.04 \AA, while the out of plane bondlengths are 
found to be same as for the case of 1\% compressed SrTiO$_3$. The in-plane angle is 
slightly reduced to 151.9$^{\circ}$ from 1\% compressed case. This results
in the same level ordering as the 1\% compressed case. 

Allowing for different magnetic configurations one finds that the ferromagnetic 
configuration is metallic while the antiferromagnetic solution is insulating. Comparing the 
energy in each case, one finds that the ferromagnetic solution has lower energy than the 
antiferromagnetic solution, though this would depend on the degree of localization. Interestingly 
as is evident from the charge density plotted for the energy interval 
from -1~eV to 0 eV, where 0 is the fermi energy,
one finds that this metallic state is strongly confined to just 
one monolayer( Fig. 6) and is in addition 
100\% spin polarized. This could have a lot of applications, one of them being in thermoelectrics as 
suggested by Ohta {\it et al}\cite{ohta}. Further the metal-insulator transition driven by a modest strain could have
applications in two state devices. The work by Verissimo-Alves {\it et al}\cite{ghos} found a spin polarized 
strongly confined metallic state in heterostructures of SrRuO$_3$ and SrTiO$_3$. Here we show that just 
one monolayer of SrO is sufficient to result in this metallic state. The competing state with an energy 
20 and 39 meV/Ru higher for the films grown on 1\% and 2\% compressed SrTiO$_3$ substrate is found to 
favor an antiferromagnetic solution. In this case, however, one finds that the d$_{xz}$ and d$_{yz}$ states are more localized. This 
drives a Jahn-Teller distortion in the system, with in-plane bondlengths now found to be equal to 1.98 and 2.0~\AA. 
As a result one finds that the down spin d$_{xz}$ orbital gets occupied at one site, while it is the d$_{yz}$ 
orbital that is found to be occupied at the neighbouring Ru site. 

{\subsection {Four monolayers of SrRuO$_3$ on SrTiO$_3$}}
Again, examining the films grown on SrTiO$_3$ without the additional strain one finds that 
while the ground states were found to be insulating at the two and three monolayers limit,
at the four monolayers limit, the system is found to be metallic.
The surface RuO$_2$ layer has a similar ordering of levels (Fig. 8 (a)-(e)) as we found at the two monolayers limit.
One finds a high spin state is realized here also, though the layer is not insulating as
we had earlier. A low density of states is found at the fermi level here.
The subsurface layer is 
found to exhibit stronger Jahn-Teller distortions than found for the three monolayer case.
As shown in the Fig. 7, the long and short in-plane Ru-O bonds are found to be 1.98 \AA~and 2.02 \AA~with the 
Ru-O-Ru angle now becoming 155$^\circ$. The reason for the more pronounced Jahn-Teller
effect is easier to understand. Unlike in the three monolayers limit, where the d$_{yz}$ and d$_{xz}$
orbitals on Ru have no states to interact with on Ti, the surface RuO$_2$ layer
provides channels for the electrons on the sub-surface d$_{yz}$, d$_{xz}$ orbitals to
delocalize. Hence there is no significant difference between exchange splitings of the 
d$_{xy}$, d$_{yz}$ and d$_{xz}$ orbitals and so the scenario found at the 
three monolayer limit doesn't happen here. 
So, as is shown in the Fig. 8(f)-(j), after the t$_{2g}$ up spin orbitals get
occupied the fourth electron goes into the d$_{xz}$
while the neighboring Ru has $d_{yz}$ occupied. However
the Jahn-Teller distortion is not large enough to make the system insulating. 

{\subsection {Magnetism at the ultrathin limit}}
In Table. II, we give the relative magnetic stabilization energies for the 
calculations corresponding to two, three and
four monolayers of SrRuO$_3$ on SrTiO$_3$. These have been given for the cases when we allowed 
GdFeO$_3$ rotations of the RuO$_6$ octahedra as well as the case when we had both 
GdFeO$_3$ rotations as well as the tilts of the octahedra. At the two monolayer limit
the system is found to be an antiferromagnetic insulator and inclusion of the tilts
changes the stabilization energy only slightly. Similar trends are seen at the three 
monolayer limit also, and the system remains to be antiferromagnetic. At the
four monolayer limit, an analysis of the density of states shows drastic differences
between the surface and the sub-surface electronic structure. The former is barely
metallic with low density of states at the fermi level and therefore favors an antiferromagnetic
arrangement of the Ru spins. The sub-surface, however has a ferromagnetic arrangement
of the Ru spins, leading to a configuration labelled as FM-AFM in Table II
and seems to be progressing towards the bulk electronic and magnetic 
structure.

In every case we have examined the dependence of U on the choice of exchange correlation
functional as well as the type of double counting scheme used. LDA calculations are
found to underestimate the distortions at the two monolayers limit.
At the two monolayers limit using LDA we found an antiferromagnetic solution for small values of U,
however one gets a ferromagnetic solution at large values of U as shown in Table III.
We also examined the role of the double counting when using
GGA+U exchange correlation functionals for both the two and three monolayers cases.
Both at the two monolayers limit and the three monolayers limit one finds a ferromagnetic
solution at lower values of U as the ground state and an antiferromagnetic solution
as the ground state at larger values of U.
For both LDA and GGA functionals, the different double counting schemes do
not have a significant effect on the results.
These results emphasize the sensitivity of the conclusions to the value of $U$.
The constrained RPA determined $U$ is able to reproduce the insulating 
ground state observed at the few monolayers limit \cite{toyota,xia} as well as
explain the exchange bias effects observed experimentally \cite{xia}.

\section {Conclusion}

We have examined the electronic structure of ultrathin films of SrRuO$_3$
grown on SrTiO$_3$. This limit turns out to be a strong playground of atomic
physics with the three monolayers becoming insulating as a result of
orbital dependent exchange splittings. At the four monolayers limit, one finds 
that the sub-surface layer which should be more delocalized than the three
monolayers limit has larger Jahn-Teller distortions, though the system
becomes metallic. Subjecting the SrRuO$_3$ overlayers to an additional 
compressive strain by straining the substrate, one finds an
insulator-metal transition at the three monolayers limit which results
in a 100\% spin polarized electron gas which is also highly confined.

\section {Acknowledgements}
KG and BM acknowledge CSIR, India for financial support.
PM thanks the Department of Science and Technology.

\begin{figure}[h]
\includegraphics[width=2.5in,angle=0]{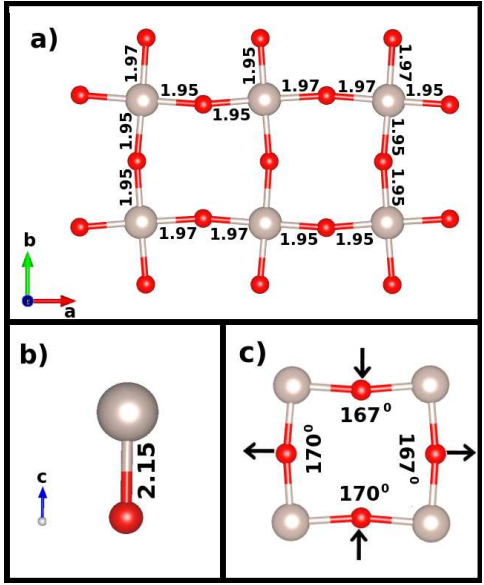}
\caption{(a) The in-plane network of Ru(large grey spheres) and O(small red spheres)
for two monolayers of SrRuO$_3$ grown on SrTiO$_3$ found in the GGA+U (U=2.5 eV, J=0.4 eV) calculations
using the Dudarev\cite{dudarev} double counting scheme. 
The Ru-O bondlengths have also been shown
in each case as well as (b) the out of plane Ru-O bondlength and (c) the Ru-O-Ru angles. The
direction of movement of the oxygen atoms are indicated by arrows.}
\end{figure}

\begin{figure}[h]
\includegraphics[width=3.0in,angle=0]{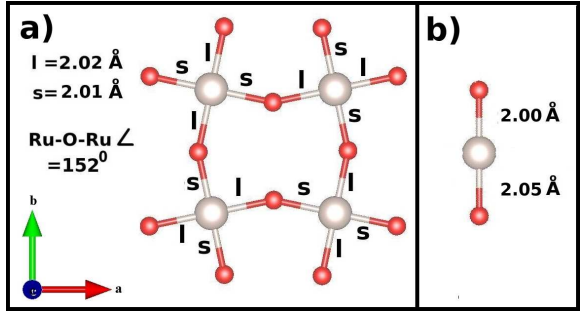}
\caption{(Color online)
The in-plane Ru (large grey spheres) and oxygen (small red spheres)
network showing Ru-O-Ru bond angles as well as Ru-O bondlengths for
three monolayers of SrRuO$_3$ films grown on SrTiO$_3$ within our GGA+U (U=2.5 eV, J=0.4 eV) calculations using
the Dudarev double counting scheme.}
\end{figure}

\begin{figure}[h]
\includegraphics[width=2.5in,angle=0]{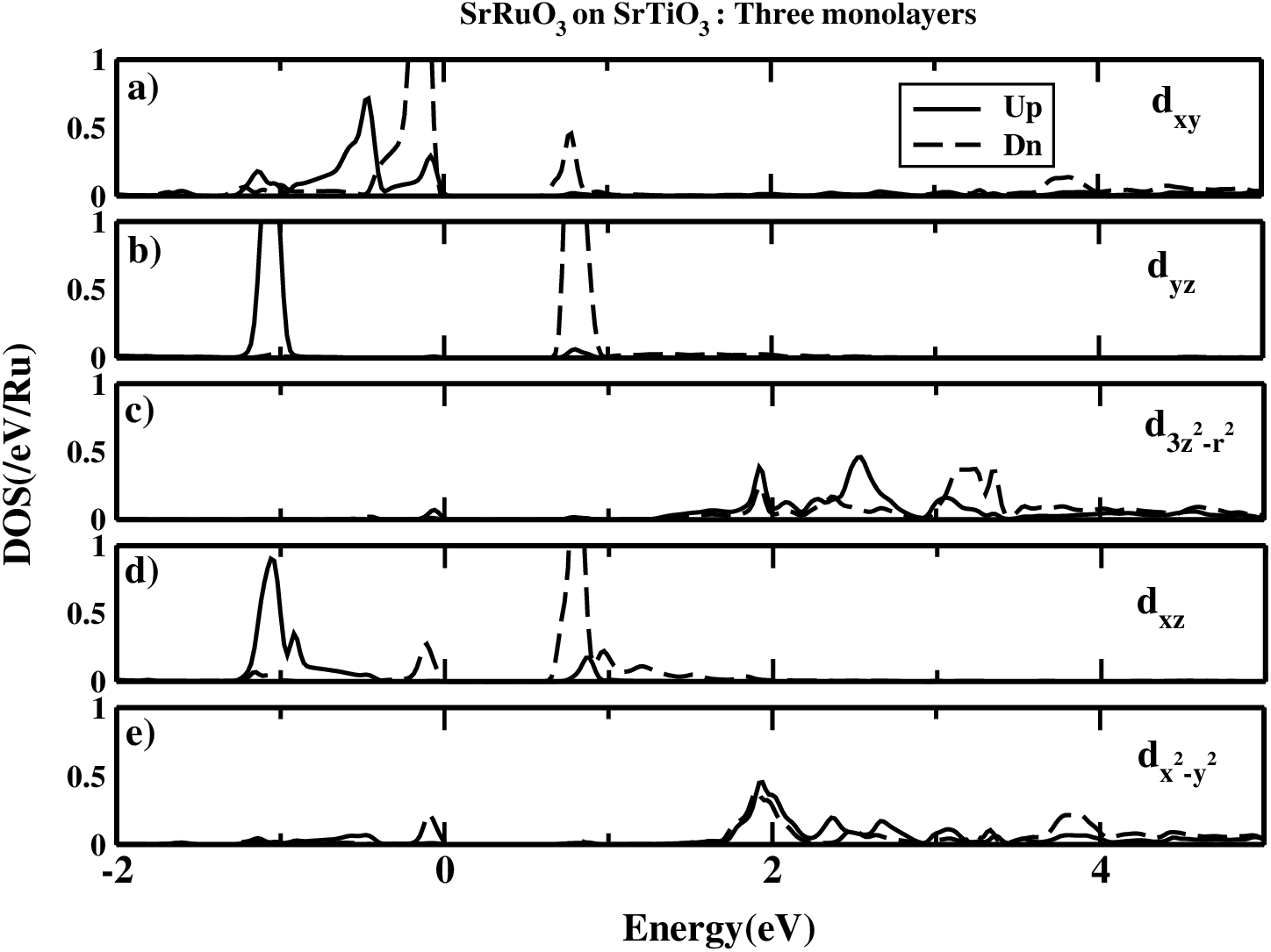}
\caption{ The up spin (solid line) as well as down spin (dashed line)
orbital projected (a)-(e) Ru $d$ partial density of states for three monolayers
of SrRuO$_3$ grown on SrTiO$_3$ using the GGA+U (U=2.5 eV, J=0.4 eV) method and the Dudarev double counting
scheme. The zero of the energy
scale is the fermi energy.}
\end{figure}

\begin{figure}[h]
\includegraphics[width=2.5in,angle=0]{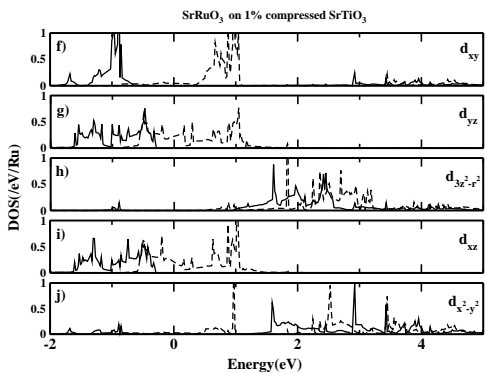}
\caption{ The up spin (solid line) as well as down spin (dashed line)
orbital projected (f)-(j) Ru $d$ partial density of states for three monolayers
of SrRuO$_3$ grown on 1\% compressed SrTiO$_3$ using the GGA+U (U=2.5 eV, J=0.4 eV) method
and the Dudarev double counting
scheme. The zero of the energy scale is the fermi energy.}
\end{figure}

\begin{figure}[h]
\includegraphics[width=2.5in,angle=0]{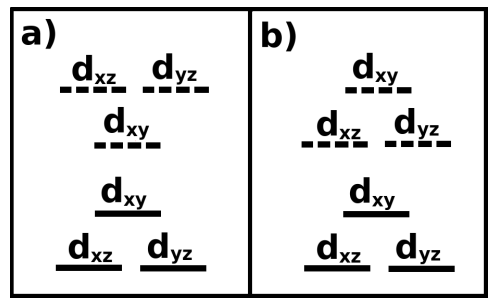}
\caption{Schematic of level ordering in up(solid) and down(dashed) spin channel
for SrRuO$_3$ on a) SrTiO$_3$ and b) 1\% compressed SrTiO$_3$ within GGA+U (U=2.5 eV, J=0.4 eV)
calculations using the Dudarev\cite{dudarev} double counting scheme.} 
\end{figure}

\begin{figure}[h]
\includegraphics[width=2.5in,angle=0]{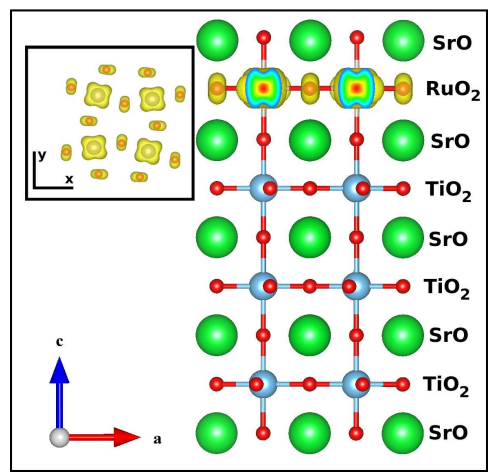}
\caption{(Color online) The layer resolved charge density in the energy 
interval -1~eV to 0~eV (fermi energy) for three monolayers of SrRuO$_3$
grown on 1\% compressed SrTiO$_3$ within GGA+U (U=2.5 eV, J=0.4 eV)
calculations using the Dudarev\cite{dudarev} double counting scheme. The view of the RuO$_2$ plane in the xy plane is shown 
in an inset.}
\end{figure}

\begin{figure}[h]
\includegraphics[width=2.5in,angle=0]{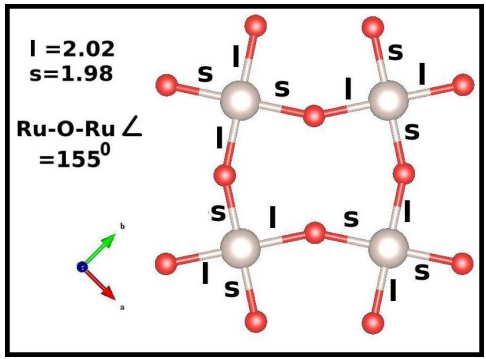}
\caption{(Color online)
The Ru (large grey spheres) and oxygen (small red spheres) network
of the sub-surface layer for four monolayers of SrRuO$_3$ grown on
SrTiO$_3$ using the GGA+U (U=2.5 eV, J=0.4 eV)
method with the Dudarev\cite{dudarev} double counting scheme. The Ru-O bondlengths as well as the Ru-O-Ru bond angles 
are shown.}
\end{figure}

\begin{figure}[h]
\includegraphics[width=2.5in,angle=0]{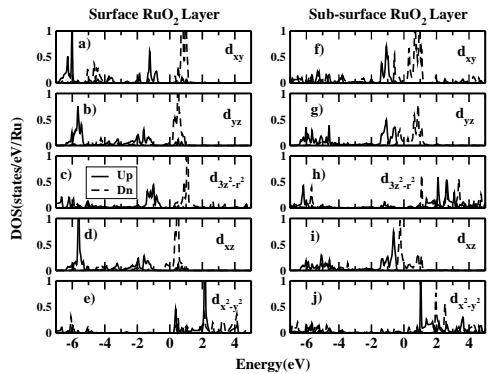}
\caption{ The up spin (solid line) as well as down spin (dashed line)
orbital projected Ru $d$ partial density of states for four monolayers
of SrRuO$_3$ grown on SrTiO$_3$ for the surface RuO$_2$ layer (a)-(e) 
as well as for the sub-surface RuO$_2$ layer (f)-(j) using the GGA+U (U=2.5 eV, J=0.4 eV)
method and the Dudarev\cite{dudarev} double counting scheme. 
The zero of the energy
scale is the fermi energy.}
\end{figure}

\begin{table}
\caption
{Ru-O bondlengths and Ru-O-Ru angles for the experimental and the 
theoretically optimized bulk SrRuO$_3$. The GGA+U\cite{ggau} method has been used for the theoretical calculations
with a U of 2.5 eV and a J of 0.4 eV.}
\begin{tabular}{|c|c|c|c|}
\hline
\hline
& \multicolumn{1}{c} {Experiment\cite{expt,expt2,expt3}} \vline & \multicolumn{1}{c} {U=2.5, J=0.4 eV} \vline\\
\hline
\multicolumn{3}{|c} {Bondlengths ($\AA$)} \vline\\
\hline
$ac$-plane&1.99/1.98 &  2.00/1.99  \\
\hline
$b$-direction&1.98 &  1.99  \\
\hline
\multicolumn{3}{|c} {Angles ($^\circ)$} \vline\\
\hline
$ac$-plane&161.1$^\circ$-162.8$^\circ$ & 158$^\circ$  \\
\hline
$b$-direction&163.1$^\circ$165.1$^\circ$ & 160$^\circ$  \\
\hline
\hline

\end{tabular}
\end{table}

\begin{table}
\caption
{Total energies in meV/Ru for all magnetic configurations for two, three and
four  monolayers of SrRuO$_3$ grown on SrTiO$_3$ 
using GGA+U exchange correlation functionals and the Dudarev\cite{dudarev} double counting scheme.}
\begin{tabular}{|c|c|c|c|c|c|}
\cline{2-4}
\hline
\hline
& \multicolumn{1}{c}{}  \vline & \multicolumn{1}{c} {GdFeO$_3$ rotations} \vline & \multicolumn{1}{c} {GdFeO$_3$+001 rotations  } \vline\\
\hline
two-mono & FM & 0 & 0  \\
         & AFM & -175 & -157  \\
\hline
three-mono   & FM & 0 & 0 \\
             & AFM & -45 & -35 \\
\hline
four-mono   & FM & 0 & 0  \\
            & AFM & -69 & -110  \\
            &FM-AFM & -82 & -129  \\

\hline\hline
\end{tabular}
\end{table}

\begin{table}
\caption
{Total energies in meV/Ru for all magnetic configurations for two and three monolayers of SrRuO$_3$ grown on SrTiO$_3$ 
with LDA/GGA for the exchange correlation functional and the double counting scheme as indicated.}
\begin{tabular}{|c|c|c|c|c|c|c|}
\hline

\multicolumn{7}{|c|}{Two monolayers} \\
\hline
& U(eV)& 0.0& 1.0& 2.0& 2.5& 3.0\\
\hline
LDA+U\cite{dudarev}        &FM& 0&  0& 0&     0& 0\\
& AFM& 8& -93& -85& -57& 348\\
\hline 
& U(eV) &0.0  &1.0 & 2.0 & 2.5 & 3.0\\
\hline
LDA+U\cite{liech}  &FM & 0 & 0 & 0 & 0 & 0\\
 &AFM & 8 & -96 & -19 & -134 & 358\\
\hline
\multicolumn{7}{|c|}{Two monolayers} \\
\hline
\multicolumn{3}{|c|}{}          &U(eV)& 1& 2.5& 4\\
\hline
\multicolumn{3}{|c|}{GGA+U\cite{liech}}       &FM &0 &0& 0\\
\multicolumn{3}{|c|}{} &AFM& 55& -568& -1168\\
\hline
\multicolumn{7}{|c|}{Three monolayers} \\
\hline
\multicolumn{3}{|c|}{}          &U(eV)& 1& 2.5& 4\\
\hline
\multicolumn{3}{|c|}{GGA+U\cite{liech}}       &FM &0 &  0 & 0\\
\multicolumn{3}{|c|}{} &AFM&303& -96& -180\\
\hline
\end{tabular}
\end{table}


\begin{references}
\bibitem{paramag}
It should be noted that these calculations are not usually
capable of capturing the paramagnetic metallic state but are very successful
in capturing magnetic ground states.
\end{references}
\end{document}